# Whistler waves in the young solar wind: statistics of amplitude and propagation direction from Parker Solar Probe Encounters 1-11

Kyung-Eun Choi[1], Oleksiy Agapitov[1], Lucas Colomban[1],
John W. Bonnell[1], Forrest Mozer[1], Richard D. Sydora[2], Nour Raouafi[3],
Thierry Dudok de Wit[2]

**ABSTRACT**

In the interplanetary space solar wind plasma, whistler waves are observed in a wide range of heliocentric distance (from ~20 solar radii (RS) to Jupiter's orbit). They are known to interact with solar wind suprathermal electrons (strahl and halo) and to regulate the solar wind heat flux through scattering the strahl electrons. We present the results of applying the technique to determine the whistler wave propagation direction to the spectral data continuously collected by the FIELDS instruments aboard Parker Solar Probe (PSP). The technique was validated based on the results obtained from burst mode magnetic and electric field waveform data collected during Encounter 1. We estimated the effective length of the PSP electric field antennas (EFI) for a variety of solar wind conditions in the whistler wave frequency range and utilized these estimates for determining the whistler wave properties during PSP Encounters 1-11. Our findings show that (1) the enhancement of the whistler wave occurrence rate and wave amplitudes observed between 25-35 RS is predominantly due to the sunward propagating whistler waves population associated with the switchback-related magnetic dips; (2) the anti-sunward or counter-propagating cases are observed at 30-40 RS; (3) between 40-50 RS, sunward and anti-sunward whistlers are observed with comparable occurrence rates; and (4) almost no sunward or counter-propagating whistlers were observed at heliocentric distances above 50 RS.

[1] Space Sciences Laboratory, University of California Berkeley, Berkeley, CA, USA
[2] Department of Physics, University of Alberta, Edmonton, Canada
[3] Applied Physics Laboratory, Johns Hopkins University, Laurel, MD, USA

# 1. INTRODUCTION

Parker Solar Probe (PSP, Fox et al. 2016; Raouafi et al. 2023) is the first spacecraft to get close enough to the Sun to sample the *in situ* characteristics of the young solar wind during its formation. PSP aims at tracing the flow of energy that heats and accelerates the solar corona and solar wind in order to address two fundamental problems in the physics of the heliosphere: coronal plasma heating and acceleration of solar wind plasmas. In November 2018 PSP became the first satellite mission to penetrate deep into the inner heliosphere, getting as close as 35.7 solar radii (RS) to the Sun during Encounter 1, and down to 12 RS after January 2022 (Fox 2016; Raouafi et al. 2023).

Solar wind electrons carry heat flux from the Sun to interplanetary space. Their distribution consists of three components: Core, Halo, and Strahl (Feldman et al. 1975; Štverák et al. 2008, 2009; Halekas et al. 2020; Salem et al. 2023). Core electrons are the thermal component, and their energy range is from 3 eV up to ~60 eV. Halo electrons are the hot component, "suprathermal" with energy ranging from ~60 eV to hundreds of eV. Strahl electrons are observed at energies across and above the halo energy region. The core and halo components are almost isotropic and close to bi-Maxwellian and bi-kappa distribution functions, respectively (Wilson et al. 2019). The strahl component has an extreme anisotropy relative to the background magnetic field direction and is characterized as a field-aligned electron beam with a narrow pitch angle $\alpha$ ($\alpha = arctan(v_\perp/v_\parallel)$) distribution (Rosenbauer et al. 1977; Feldman et al. 1975). In the expanding solar wind, the solar wind magnetic field strength becomes weaker, and because of the constraint of kinetic electron energy and the conservation of the first adiabatic invariant, $v_\perp$ should decrease. However, the beam angular width of strahl observed at 1 a.u. is often not as narrow as expected (e.g., Feldman et al. 1978; Lemons and Feldman 1983; Pilipp et al. 1987; Berčič et al. 2019), which is presumably due to scattering processes caused by Coulomb collisions (however, Lemons and Feldman (1983) have shown that collisions are not always sufficient to describe the observed strahl width, especially at lower energies) and/or interaction with waves (e.g.,

Vocks et al. 2005; Pagel et al. 2007; Colomban et al. 2024). The evolution of the electron distribution function with heliocentric distance was reported in several previous studies (e.g., Maksimovic et al. 2005; Štverák et al. 2009; Graham et al. 2017). Graham et al. (2017) found that beyond 1 a.u. the strahl population is likely to be completely scattered and presumably contributes to the halo during propagation from 1 a.u. to approximately 8 a.u.

The first very large amplitude whistler in the radiation belt was reported measured with STEREO by Cattell et al. (2008). However, because STEREO only measured electric fields, it was limited in its ability to estimate the magnetic fields of waves and wave normal angles (WNAs). Later, studies using Wind (e.g., Wilson et al. 2011), in which magnetic field measurements are available to investigate whistler waves, showed that most (30/46) of the whistler waves propagate within 20° relative to the background magnetic field. The recent studies of whistler waves in the solar wind based on PSP measurements by Jagalarmudi et al. (2020), Kretzschmar et al. (2021), Cattell et al. (2022), Froment et al. (2023), and Colomban et al. (2023) have shown the occurrence rate and amplitude of whistler waves observed in the solar wind depend on the heliocentric distance and this raises the question about the mechanism(s) of their local generation. Quasi-parallel whistler waves are dominant in observations reported by (Stansby et al. 2016; Tong et al. 2019; Kretzschmar et al. 2021; Froment et al. 2023; Colomban et al. 2023, 2024) in the solar wind. Whistler heat-flux instability, induced by an electron thermal beam can produce quasi-parallel whistler waves (Gary et al. 1975; Vasko et al. 2020). Quasi-parallel whistlers can also be generated by electron cyclotron resonance instability guided by anisotropic electron distributions (Agapitov et al. 2020; Artemyev et al. 2013; Feldman et al. 1976; Gary et al. 1975, 1994; Lee et al. 2018, 2019; López et al. 2019; Karbashewski et al. 2023; Roberg-Clark et al. 2018, 2019; Shaaban et al. 2019; Tong et al. 2019; Vasko et al. 2020; Zenteno-Quinteros & Moya 2022), or driven by the relaxation of the sunward electron deficit (Berčič et al. 2021). Kretzschmar et al. (2021) reported that at 0.5 - 1.0 a.u., all whistler waves observed by Solar Orbiter propagate anti-sunward. Karbashewski et al. (2023)

discovered the counter-propagating whistlers at ~41 RS co-located with magnetic dips and/or sharp rotations of the solar wind magnetic field direction associated with the magnetic switchbacks – the solar wind structures with rapid deflections of the magnetic field direction inside (Mozer et al. 2020; Krasnoslelskikh et al. 2020; Larosa et al. 2021; Agapitov et al. 2023). The boundaries of these structures usually have dips in the magnetic magnitude (Farrell et al., 2020, 2021; Froment et al. 2021) that naturally evolved during the generation (Drake et al. 2021) and evolution (Agapitov et al. 2022). Karbashewski et al. (2023) performed a numerical estimation of the growth rate and showed that the cyclotron instability can be a generation mechanism(s) for the observed counter-propagating whistler waves.

The effects of interaction between electrons and whistler waves were reported based on the Wind (e.g., Moullard et al. 2001; Vocks et al. 2005; Vocks & Mann 2003, Willson et al. 2013), Cluster (e.g., Kajdič et al. 2016), and STEREO (e.g., Cattel et al. 2020) observations. Numerical models have been implemented to examine the effects of wave-particle interactions in the solar wind, demonstrating the scattering of strahl and the formation of halos (Pierrard et al. 2011; Roberg-Clark et al. 2019; Saito & Gary 2007; Seough et al. 2015; Tang et al. 2020; Vo et al. 2022; Vocks 2012; Vocks et al. 2005).

Halekas et al. (2020, 2022) and Abraham et al. (2022) reported the electron distribution function features in the young solar wind from the PSP observations. Halekas et al. (2020, 2022) reported that the strahl electrons observed by PSP were narrower than expected and dominant in suprathermal populations, while the halo nearly disappeared near perihelion, much smaller even than those at 0.3 a.u. Their results showed that the strahl and halo electrons were not consistent with the radial scaling expectation based only on the divergence of the interplanetary magnetic field producing strahl populations. In contrast, the observed core population of electrons near perihelion tends to follow the radial scaling expectation. As a result, these observations of suprathermal populations in the solar wind may show the exchange of halo and strahl electrons by pitch-angle scattering caused by wave-particle interactions (Agapitov et al. 2020;

Cattell et al. 2021b; Froment et al. 2023; Jagarlamudi et al. 2021; Karbashewski et al. 2023; Malaspina et al. 2020; Mozer et al. 2020). For quasi-parallel whistler waves, the propagation direction relative to the strahl speed determines the efficiency of wave-particle interaction. The derivation of the wave propagation direction and statistical mapping of amplitudes and occurrence rates of sunward propagating whistlers (SWPW) and anti-sunward propagating whistlers (A-SWPW) waves is needed for developing realistic models of wave-particle interactions and elucidating the potential generation mechanisms of these waves. In this paper, we apply an analysis technique and the statistics of whistler waves, based on the PSP measurement of the wave electric to magnetic field ratio to reveal the direction of propagation of whistler waves from the spectral power of magnetic ($B_w$) and electric ($E_w$) field continuously collected by PSP (one component of $E_w$ perturbations and three components of $B_w$ (two components of $B_w$ after February 2019)). We have obtained the whistler waves' amplitude and propagation direction (sunward or anti-sunward), statistically, applying this technique to the data collected by PSP during Encounters 1-11. The data description and the details of the processing technique along with examples of the application are shown in Section 2. The effective antenna length $L_{eff}$, which we have used in this work was determined based on the whistler wave dispersion relation by making use the technique proposed by Karbashewski et al. (2023). Section 3 presents the statistics of whistler wave properties from the first eleven PSP encounters. Section 4 contains a short summary of the main results.

## 2. DATA DESCRIPTION AND METHODOLOGY

### 2.1. Parker Solar Probe data description

In this paper, we use the magnetic and electric field spectrogram data from the PSP Search Coil Magnetometer (SCM, Dudok de Wit et al. 2022) and the PSP Electric Field Instrument (EFI, Mozer et al. 2020), which are part of the FIELDS suite (Bale et al. 2016) on board the Parker Solar Probe. Data are preprocessed on board PSP by the Digital Filter Board (DFB, Malaspina et al. 2016) which provides the spectrograms of magnetic and electric field perturbations in 54 frequency channels. The background solar wind plasma parameters were estimated based on the measurements of DC magnetic field by the fluxgate magnetometer (MAG); the moments of proton distribution function - from the Solar Probe Cup (SPC, Case et al. 2020) and Solar Wind Electrons Alphas and Protons (SWEAP, Kasper et al. 2016). The EFI and SCM spectrogram data have time resolution of ~28 seconds (i.e., averaging ~28 seconds of onboard fast Fourier transform data) and a frequency range from 22 Hz to 4.5 kHz which covers the whistler wave frequency range in the solar wind (typically ranging from 100 Hz to 300 Hz (Cattell et al. 2021b; Froment et al. 2023; Karbashewski et al. 2023)). Dudok de Wit et al. (2022) and Froment et al. (2023) demonstrated the consistency of the parameters of whistler waves (wave spectral power and polarization parameters) derived from the spectrogram data and from the high-sampling burst mode waveform data (150 kS/s).

We apply the whistler dispersion to the spectral data to derive the SWPW or A-SWPW waves by making use of the Ew/Bw ratio and the bulk solar wind speed (Karbashewski et al. 2023). To illustrate the application of the methodology and validate using of the PSP spectral data, we focus on a 1-hour interval from 10:00 to 11:00 UT on November 3, 2018 (previously discussed by Cattell et al. (2021b) and Karbashewski et al. (2023)). This interval contains multiple bursts of whistler waves including sunward, anti-sunward, and counter-propagating waves (Karbashewski et al. 2023). Figure 1 shows an overview of the interval: the background interplanetary magnetic field, solar wind bulk velocity and proton density, and spectral power in the whistler frequency range with multiple bursts of whistler waves. The radial component (denoted by red) of the background magnetic field $\vec{B}_{SW}$ was sunward with significant tangential (denoted by blue) and normal (denoted by green) components (in the RTN coordinate system, Figure 1a), while the bulk velocity $\vec{V}_{SW}$ was predominantly radial (Figure 1b). Figure 1d and Figure 1e present the spectral power of magnetic field $B_w$ and electric field $E_w$ perturbations. While $B_w$ is the sum of all three components, the power spectrum of $E_w$ is derived from antennas 1 & 2.

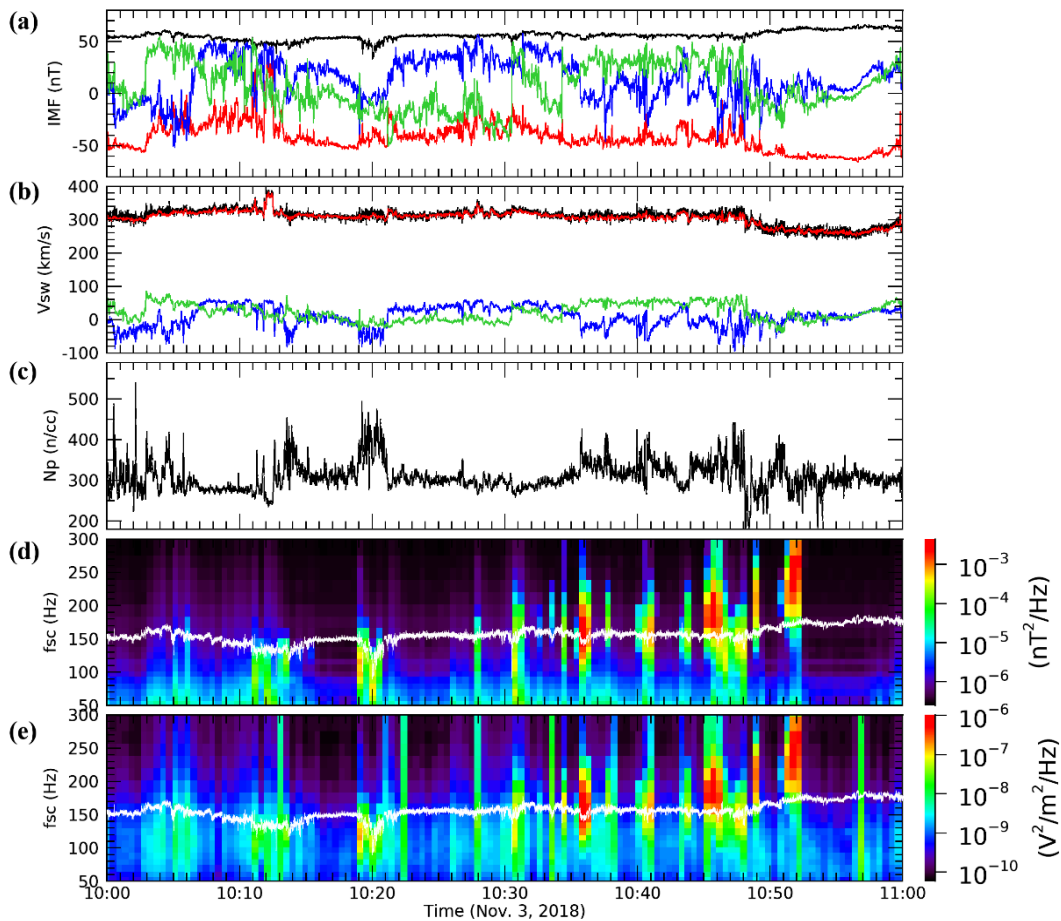

**Figure 1.** One-hour interval with multiple whistler wave events from 10 to 11 UT on November 3, 2018 collected by Parker Solar Probe. (a)-(c) interplanetary magnetic field (MAG data), solar wind proton bulk velocity and proton density (SWEAP data). The magnitude of quantities and each component in the RTN coordinate system are presented in different colors, black, red, blue, and green, respectively. (d)&(e) Power spectral density (PSD) of the magnetic and electric field perturbations in the spacecraft frame frequency ($f_{sc}$). The white curves indicate local $0.1 f_{ce}$.

In the case of a plasma frame moving with velocity $\vec{v}$, where $|\vec{v}| << c$, the electromagnetic field $\vec{E}_w$ and $\vec{B}_w$ measured by the non-moving observer (indicated by unprimed) is related to the plasma frame field $\vec{E}_w'$ and $\vec{B}_w'$ via the Lorentz transform:

$$\vec{B}_w = \vec{B}_w' \qquad (1)$$

$$\vec{E}_w = \vec{E}_w' - \vec{v} \times \vec{B}_w \qquad (2)$$

In the solar wind the $E_w/B_w$ ratio derived from spectral power measured by PSP depends on the wave propagation direction and the solar wind bulk velocity ($V_{sw}$): $E_w/B_w = |\vec{E}'_w|/|\vec{B}'_w| \pm V_{sw}$. This allows one to determine the direction of propagation of the observed whistler waves taking into account that $V_{sw}$ is about 0.2-0.8 of $V_{ph}$ at heliocentric distances of 25-50 RS (as shown in Figure 2a).

The dynamics of the $E_w/B_w$ ratio for the interval from figure 1 is presented in Figure 2 (the saturated electric field data, shown in Figure 1e, is removed in all frequency ranges). The $E_w/B_w$ ratio depends on wave frequencies and represents a proxy for the phase velocity (in the outward solar wind flow at the corresponding frequency in the spacecraft frame). Following the results of (Froment at al. 2023; Karbashewski et al. 2023), we assume waves propagating along the background magnetic field $\vec{B}_{SW}$ inward (sunward propagating whistler wave, SWPW) or outward (anti-sunward propagating whistler wave, A-SWPW). The results presented in Figure 2 show that the $E_w/B_w$ changes about twice corresponding to wave propagation direction (blue and red colored dots in Figure 2b indicate A-SWPW and SWPW, respectively). The feature of this particular interval is that lower values of $E_w/B_w$ correspond to higher wave spectral power (the black curve in Figure 2b), reflecting sunward propagation of high amplitude whistler waves, which is consistent with the results of Karbashewski et al. (2023).

Figure 2c shows $v_{ph}$ in the plasma frame from the cold plasma dispersion (the black curve) for parallel propagation. The effects of the moving plasma media are shown with the red dashed curve (SWPW) and the blue dashed curve (A-SWPW), which are calculated in the SC-frame using the averaged parameters derived from the interval shown in Figure 1. As an example of the significant Doppler shift, the measurements at $f_{SC} = 123.6$ Hz correspond to 184 Hz (SWPW, +48%) and 84 Hz (A-SWPW, -32%) respectively in the plasma frame (more details of Doppler shift described in Section 2.3). Then, to compare with observed $E_w/B_w$ ratio values, these theoretical phase speeds must be considered by their effective length (in detail in following Section 2.2) indicated by two solid lines.

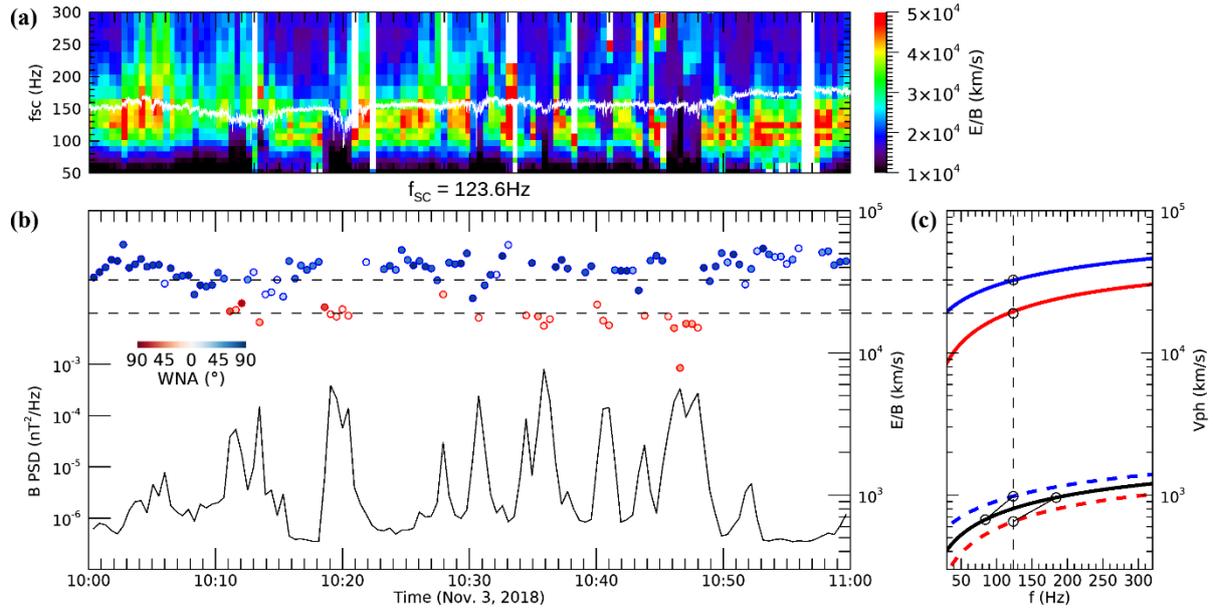

**Figure 2.** (a) The ratio of the electric field to the magnetic field $E_w/B_w$ obtained from spectral data. (b) The measured $E_w/B_w$ values (the red dots indicate sunward whistlers and enhanced $B_w$ values) and PSD of magnetic field $B_w$ data (black line) at $f_{SC} = 123.6$ Hz. Colors of circles denote WNA values as shown in the color bar. (c) Theoretical phase speed and Doppler shifted cases of A-SWPWs and SWPWs. The black curve represents the phase speed of the whistler wave calculated from the dispersion relation. Blue and red dashed curves show those in the SC-frame for A-SWPWs and SWPWs, respectively. Blue and red solid curves represent the effective lengths applied corresponding to the frequency of 123.6 Hz.

## 2.2. Effective length of the PSP EFI antennas ($L_{eff}$)

The calibration of the PSP EFI electric field measurements, i.e., evaluation of the antenna's effective length ($L_{eff}$) has been done in the whistler frequency range (~50 Hz to 500 Hz) based on the whistler dispersion and sunward/anti-sunward relative difference in $E_w/B_w$. $L_{eff}$ can exceed the physical size and separation of the antennas, leading to overestimation of $E_w$ (Agapitov et al. 2020; Karbashewski et al. 2023). Thus, continuous calibration of electric field measurements is critically needed for correct wave processing.

Figure 3a shows a scatter plot of $E_w$ and $B_w$ collected at $f_{SC}$ = 123.6 Hz, revealing two distinct populations at different slopes of $E_w/B_w$ representing $v_{ph}(f_{SC})$ of A-SWPW (blue) and SWPW (red). Figure 3b represents the electric field antennas effective length, $L_{eff}$, published by Karbashewski et al. (2023) for 1-hour interval. The application of $L_{eff}$ results in adjustments to the measured $v_{ph}(f_{SC})$ values due to a significantly overestimated electric field. The theoretical values are shown with the dashed blue line (A-SWPW) and the dashed red line (SWPW) at the bottom right in Figure 3a. Two solid lines represent these theoretical lines following the application of $L_{eff}(f_{SC} = 123.6\,\text{Hz})$, relative to physical antenna length (3.5 m), indicated by multiplying the value by a factor of $L_{eff}$, relative to the electric field.

In Figure 3(b), the effective lengths at 123.6 Hz are 29.2, 31.6, and 27.2 m, measured under different conditions: high wave amplitude using burst mode data (by 2-d histogram of the number of data points in the background), the entire 1-hour spectral data using best fit (a star), and low wave amplitude using spectral data (by error bars), respectively. During this 1-hour period, the plasma conditions with $f_{ec}/f_{pe} \approx 100$ ensure that only whistler waves can exist, even at low wave amplitudes comparable to noise levels. The effective length for the low amplitude interval, shown by the error bars, exhibits a consistent shape and comparable values to those observed during high amplitude whistler events (burst mode data), particularly when those noises are assumed to be quite oblique WNAs (>50°) and are then reflected as a parallel WNA by theoretical estimation (denoted by boxes of error bars). These WNAs of noises are obtained by cross-spectra data for 1-hour interval. The weight coefficient between the high wave amplitude data from burst mode and the low amplitude from spectral data is 1.32, within the frequency range of 50 to 300 Hz (in the spacecraft frame) over this 1-hour period (not shown in here).

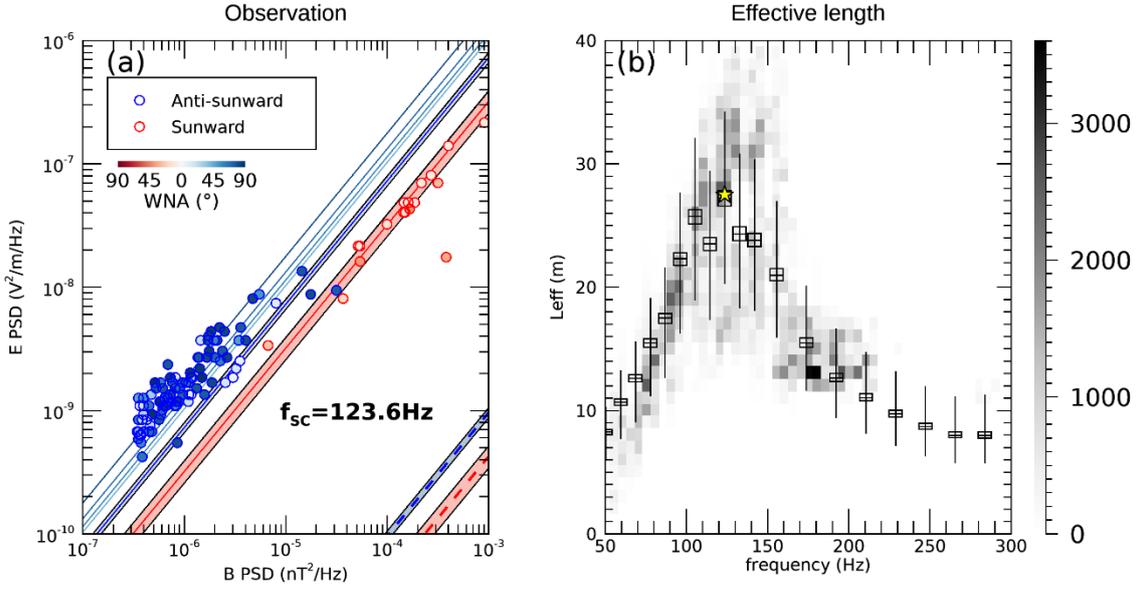

**Figure 3.** Measured E/B ratio and the utilization of effective length for 1-hour period. (a) Comparison of the measured (circles) and expected (lines) $E_w/B_w$ ratio for 1-hour at 123.6Hz in the SC-frame. Colors of circles indicate measured WNAs and colored thin lines are theoretical $E_w/B_w$ values for oblique WNAs. (b) The effective length, $L_{eff}$, as published by Karbashewski et al. (2023). Color bar show the number of data points. The error bars represent the median values and standard deviations of $L_{eff}$ obtained from spectral data including the cases of parallel WNA (WNA=0°) calculated from different WNAs by applying theoretical electric fields.

In this work, we used our technique for estimation of $L_{eff}(\omega)$ in the whistler frequency band and for a wide range of plasma densities. $L_{eff}(f_{SC})$ for a given frequency in the SC-frame can be estimated from whistler waves measured by PSP comparing the apparent phase velocity $E_w/B_w$ and expected phase velocity $v_{pl}$:

$$L_{eff}(f_{SC}, N_p) = L_{antenna} \frac{|E_w(f_{SC})|/|B_w(f_{SC})|}{v_{ex}(f_{SC}, N_P)} = L_{antenna} \frac{|E_w(f_{SC})|/|B_w(f_{SC})|}{v_{ph}(f_{pl}, N_P) + \vec{V}_{SW} \cdot \hat{k}(f_{pl})}, \quad (3)$$

applying the apparent velocity from the ratio of the measured $E_w$ and $B_w$ spectra, the solar wind flow speed $\vec{V}_{SW}$, the unit wavevector $\vec{k}/|k|$, and the theoretical $v_{ph}(f_{pl}, N_p)$. Here WNA of whistler waves have been assumed as quasi-parallel to the background magnetic field.

Since Encounter 8, PSP has provided solar wind parameters below 20 RS for the first time and allowed us to investigate $L_{eff}$ in a dense plasma. In the measurement of effective length at high densities (over ~700 cm$^{-3}$), more than half the cases are below 35 RS, and the frequency range above 100 Hz could not be investigated due to the lower hybrid waves with frequency shown in Figure A1. As a result, the standard error is higher than 1 m (denoted by white curves in Figure 4b).

Although not shown here, we tested the effect of other solar wind parameters which are speculated to affect the electric field antenna response well, such as the local plasma density, temperature, and electron cyclotron frequency. The density was the most relevant parameter to the effective length at a given frequency. As we mentioned above, the wide range of densities measured from PSP observations provides the opportunity to estimate the effective length covering a variety of solar wind conditions. The effective length along the $L_{antenna}/\lambda_D$, where $L_{antenna}$ = 3.5 m is the physical length of antenna and $\lambda_D = \sqrt{\frac{\epsilon k_B T}{nq^2}}$ (using $T_e = 30 eV$), is shown in Figure 4.

In Figure 4, we present a comprehensive map of the effective length $L_{eff}(f_{SC}, N_p)$ of the PSP electric field antenna. To validate our results, we compare $L_{eff}$ specific density regions, given in Figure 4a, to those in the case study by Karbashewski et al. (2023). Figure 4a shows that $L_{eff}$ is in the same frequency range as in Figure 10 of (Karbashewski et al. 2023), where the density was ~290 cm$^{-3}$ and has the same frequency dependence varying from ~14 m to ~22 m. It is comparable and slightly longer than those presented in (Karbashewski et al. 2023). Above 150 Hz, $L_{eff}$ decreases to about 12 m at 190 Hz before appearing at the plateau around 9 m below 300 Hz. Further higher frequency ranges have similar $L_{eff}$ of approximately 7.5 m. Overall, $L_{eff}(f_{SC}, N_p)$ for the density of 290 cm$^{-3}$ ranges from ~6 m to ~22 m for the frequency range 50-600 Hz with the longest $L_{eff}(f_{SC}, N_p)$ between 100-150Hz, in this study.

Figure 4b displays the median quantity of $L_{eff}$ versus frequency at the same density bin. The frequency bin size follows that of the DFB spectral data products. In density, a bigger bin size has been applied at high density to accommodate the smaller number of data points. Additionally, if the angle difference between solar wind bulk flow and the magnetic field is large (more details in Section 2.3.), the topology of the antenna sensor in response to wave perturbations may lead to uncertainties in the electric field measurements as we describe in Section 2.2. To reduce the impact of this effect on our analysis, we provide the effective length map where the $\alpha_{BV}$ is less than or equal to 30 degrees. As a result, we have obtained a comprehensive effective length map $L_{eff}(\omega, N_p)$ for PSP observations over the whistler wave mode frequency range. $L_{eff}$ values cover 2.1 to 36 m within the confidence interval (indicated by the white curve on the map), about ten times larger than $L_{antenna}$ and tend to be proportional to the density and inversely proportional to the frequency. It implies that the electric field measurements are sensitive to the local solar wind conditions.

$L_{eff}$ decreases with frequency except for a particular density range. This feature occurs at densities of ~250 to ~350 cm$^{-3}$ ($L_{antenna}/\lambda_D \sim 1.5$ where $T_e = 30 eV$) and frequencies below 150 Hz. Figure 4c shows how much the $L_{eff}$ values differ according to density at the same frequency channel. One of the highest $L_{eff}$ values in Figure 4a is at 123.6 Hz. For this specific frequency, $L_{eff}$ varies from ~3.7 to ~22.7 m depending on the density conditions. Near the density value 300 cm$^{-3}$ ($L_{antenna}/\lambda_D \sim 1.5$ where $T_e = 30 eV$) it shows the largest effective length and over 250 cm$^{-3}$ ($L_{antenna}/\lambda_D > 1.35$ where $T_e = 30 eV$) $L_{eff}$ values are similar, between 15 to 23 m (with large error bars over 500 cm$^{-3}$ ($L_{antenna}/\lambda_D > 2$)). One can see that this density range corresponds to the case where the physical length of the antenna is greater than the Debye length ($L_{antenna}/\lambda_D > 1$). Two vertical dotted lines denote different cases for $L_{antenna}/\lambda_D = 1$, when $T_e$ is 20 eV (orange color) or 40 eV (magenta color), respectively. It implies that stable measurements occur where the antenna length is over several Debye lengths. Through this result, we find that different plasma contributions from various solar wind regimes make $L_{eff}$ differ even in the same frequency range. Consequently, by applying our technique to this result of a comprehensive effective length map (Figure 4b), the propagation directions of whistler waves from Encounter 1-11 of PSP measurements have been determined in this work and we present those statistics in Section 3.

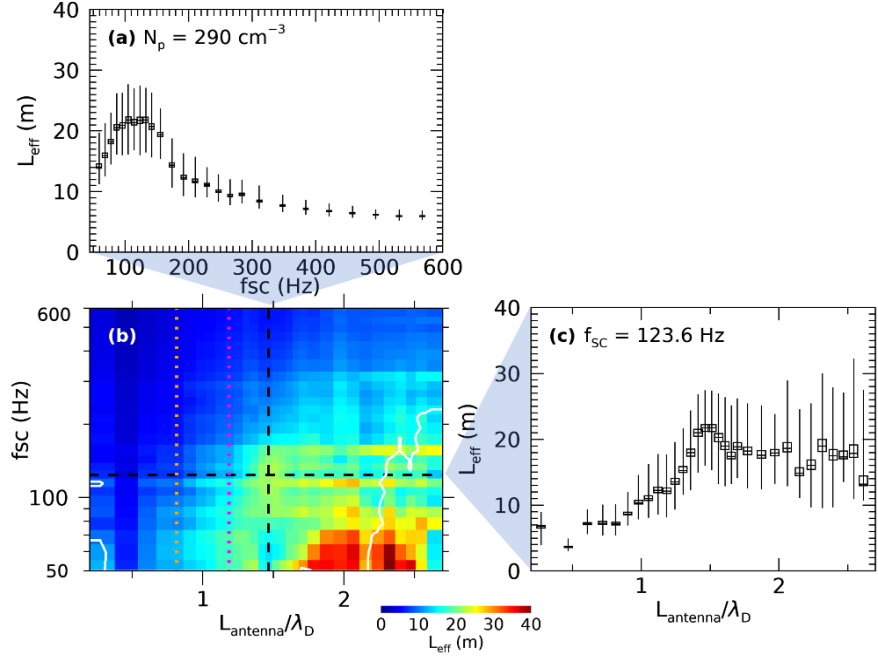

**Figure 4.** Effective electric antenna length ($L_{eff}$) on the $L_{antenna}/\lambda_D$ ($where\ T_e = 30eV$) and wave frequency ($f_{SC}$) in the SC-frame for the PSP Encounters 1-11. (a) $L_{eff}$ for $N_p = 290$ cm$^{-3}$ (indicated by a dashed vertical line in panel (b)). (b) - $L_{eff}$. (c) - $L_{eff}$ on $L_{antenna}/\lambda_D$ for $f_{SC} = 123.6$ Hz (indicated by a dashed horizontal line in panel (b)).

The $L_{eff}$ map in Figure 4 facilitates the accurate estimation of the electric fields measured by EFI antennas in the whistler wave frequency range. Steinvall et al. (2021) presented the effective length of the Solar Orbiter's electric field measurement, in which a constant electron temperature is used. We note that the effective lengths show a unique pattern in a particular density range (250-350 cm$^{-3}$) which seems to have originated from other effects but appear to be unrelated to the electron temperatures (Romeo et al. 2023).

**2.3. Analysis technique to determine propagation directions of whistler waves**

In the SC-frame, $V_{SW}$ produces a significant Doppler shift: a SWPW (A-SWPW) wave is observed at a down-shifted (up-shifted) frequency, $f_{SC}$. Schematically the geometry of field measurements onboard PSP in the solar wind flow ($\vec{V}_{SW}$) is shown in Figure 5a. $\alpha_{BV}$, the angle between $\vec{B}_{SW}$ and $\vec{V}_{SW}$ (denoted by cyan color in Figure 5a and measured data in Figure 5b), can be used as an approximation of the angle between $\vec{V}_{SW}$ and whistler wave normal because WNA is predominantly quasi-parallel with respect to the background magnetic field (Froment et al. 2023).

The wave parameters in the plasma frame can be evaluated from Eqns.1 and 2 by making use of the whistler dispersion relation:

$$v_{ph}(f_{SC}) = v_{ph}(f_{pl}) + \vec{V}_{SW} \cdot \hat{k}(f_{pl}) = v_{ph}(f_{pl}) + V_{SW} |\hat{k}(f_{pl})| \cos(\alpha_{BV})$$
(4)

Here, $v_{ph}(f_{pl})$ is calculated from the cold plasma dispersion relation in the plasma frame. The additional phase velocity from the outflow speed is applied by $\vec{V}_{SW} \cdot \hat{k}(f_{pl})$, and $\hat{k}(f_{pl})$ is assumed to be field-aligned.

However, the spectral data only include a single component of the electric field measured from antennas 1 and 2. This impacts the estimate of phase velocity in two ways: first, if the local magnetic field lies in or near the plane of the antennas, only one of the two transverse $\vec{E}_w$ directions is measured, leading to statistical underestimates of and uncertainties in the $|\vec{E}_w|$ magnitude; second, even if $\vec{B}_{SW}$ is radial, i.e. normal to the antenna plane, only one of the two transverse components of $\vec{E}_w$ is measured, again leading to underestimation of the magnitude of $|\vec{E}_w|$, and $V_{ph}$. The first effect can be minimized by considering intervals where local $\alpha_{BV}$ is small, which for our study we limit to 30 degrees ($\alpha_{BV} < 30°$ or $\alpha_{BV} > 150°$) for evaluating the antenna effective length. For the processing of the propagation direction this leads to spread of $E_w/B_w$ in the distribution between the theoretical value of $E_w/B_w$ and $(E_w^2/2B_w^2 + V_{SW}^2)^{1/2}$, which however, allows us to distinguish between SWPW and A-SWPW in a wide range of $\alpha_{BV}$. The second effect can be taken into account by applying a factor of $\sqrt{2}$. This effect is also influenced by the WNA of the wave, but the fractional deviation is only around 0.02 for WNAs below 20 degrees, which typical WNAs of whistler waves are in the $0.1f_{ce}$ to $0.5f_{ce}$ frequency range (Froment et al. 2023).

Figure 5c shows $E_w/B_w$ in the $\alpha_{BV}$ domain. The two black solid lines represent the theoretical phase velocity from $\alpha_{BV} = 0°$ to $90°$ for A-SWPW (upper line) and SWPW (lower line), respectively. On the other hand, there are geometrical limitations of measuring $\vec{E}_w$ with only one component. Therefore, the two dashed lines indicate the bottom limit of the distribution phase velocity value, which decrease at $\alpha_{BV} = 90°$ with solar wind bulk speed (taking into account the factor of $L_{eff}$). These lines are based on an average of 1-hour intervals to represent all data points, but the processing of each data point is based on the local solar wind conditions (marked colors of filled circles in the same legend in Figure 3a). The circle size is proportional to $B_w$. We excluded A-SWPWs detected below the lower hybrid frequencies (in the plasma frame). While high-amplitude whistler waves focus on the parallel WNAs (black lines), low-amplitude points are distributed broadly to oblique WNAs. Note that here we do not determine WNAs with the techniques in this paper.

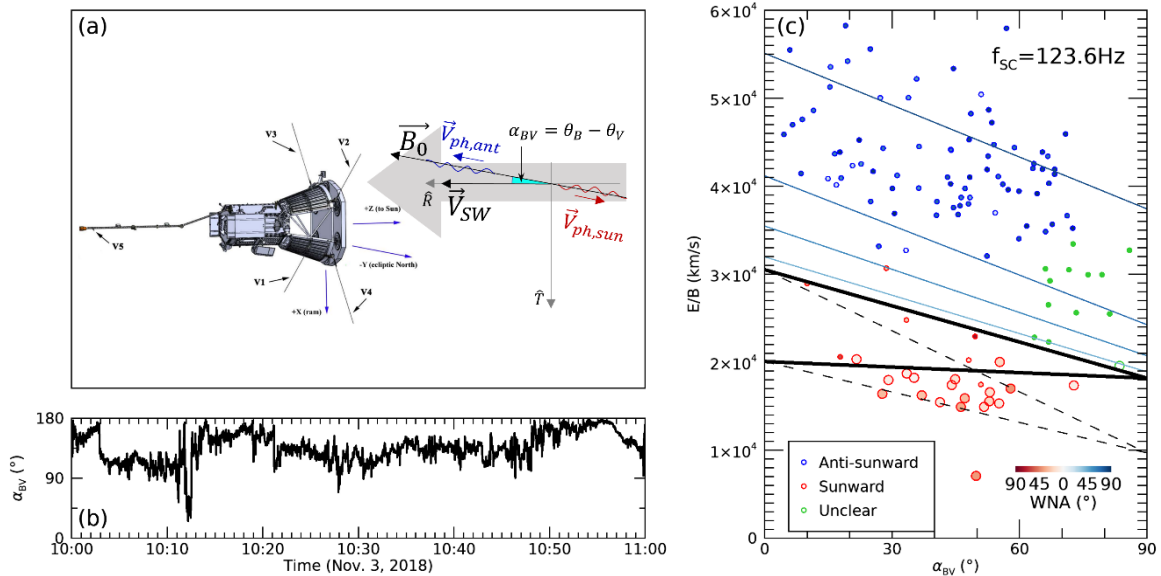

**Figure 5.** Ultimate Whistler wave's direction determination for 1-hour interval incorporating the topology of the background magnetic field (a) PSP position relative to the radial direction to the Sun (adapted from Malaspina et al. (2016)) and schematic of $\vec{B}_{SW}$ (or $\vec{B}_0$) and $\vec{V}_{SW}$ and its implications on the electric field measurements inferred from the direction of wave propagation. (b) Dynamics of the angle between the background magnetic field and flow velocity in the solar wind. (c) Distribution of measured $E_w/B_w$ ratio ($f_{SC} = 123.6$Hz). The two black solid lines represent the theoretical $E_w/B_w$ values, for A-SWPWs and SWPWs, respectively, based on the average of local solar wind parameters over a 1-hour period. The two dashed lines mark where the electric field magnitude is poorly estimated due to the magnetic field lying in the plane of the antennas. Colored lines are theoretical $E_w/B_w$ values for oblique WNAs (corresponding angles in degrees on the color bar).

## 2.4. Validation

Figure 6 is the result of determining the propagation direction of whistler waves on the interval form Figure 1. Four whistler waves events (Figures 6a-6d) were recorded in burst mode and previously presented by Karbashewski et al. (2023); they are indicated by black arrows in Figure 6e. Figures 6a-6d represent the individual Poynting fluxes obtained from the burst waveform data using wavelet transform and singular value decomposition. These Poynting vectors, were used for validation of the analysis technique described above. Consequently, the results obtained by making use of the waveform data are perfectly matched with those of our results shown in Figure 6e. Figure 6e presents a map of the whistler wave propagation direction for 1 hour obtained by making use of our technique. Magnetic dips (implied by the $0.1\,f_{ce}$ line) at around ~10:10UT, ~10:20UT, and ~10:30UT are associated with the enhanced level of SWPW activity.

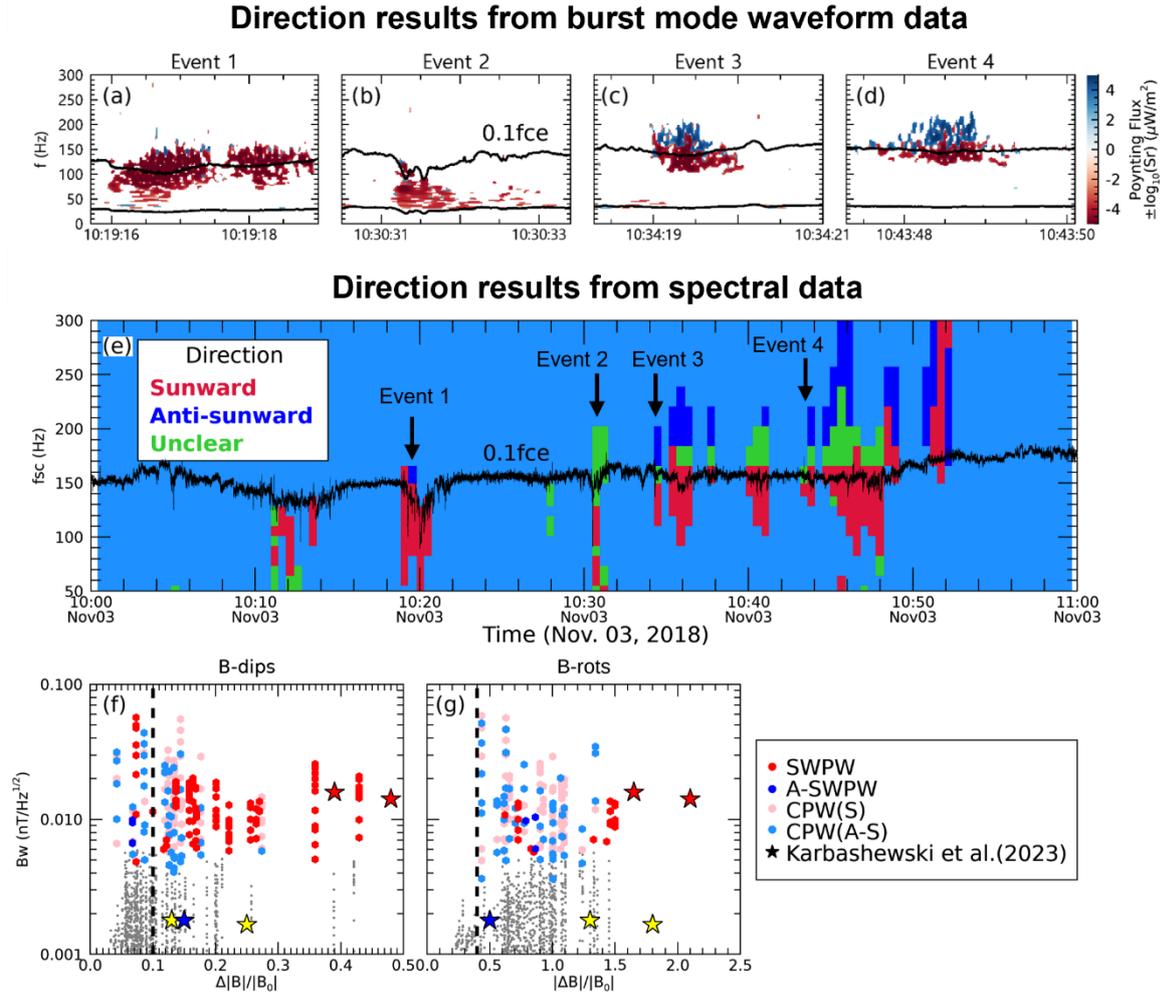

**Figure 6.** (a)-(d) Each panel presents Poynting flux for different whistler waves corresponding to four arrows in panel(e). (a) The burst waveform starting at 10:19:15.56 UTC on November 3, 2018. (b) starting at 10:30:30.15 UTC. (c) starting at 10:34:18.22 UTC. (d) starting at 10:43:47.29 UTC. Two black curves indicate $0.1 f_{ce}$ and $f_{LH}$. (e) the results of determining the propagation direction in time and frequency. The black curve represents $0.1 f_{ce}$. Arrows indicate whistler waves validated by burst waveform data. (f)&(g) show the distribution of whistler waves amplitude on variation of the magnetic field relative to background fields. (f) the dip of magnetic field magnitude and (g) the rotation of magnetic field components.

Figure 6f presents the $\Delta|B|/|B_0|$ (meaning the decrease in local magnetic field magnitude relative to the background magnetic field) and the $B_w$ wave amplitude for 1 hour (with solar wind outflows denoted by gray color). In this 1-hour interval, most A-SWPWs (except only for a few events shown in blue in Figures 6f and 6g) are accompanied by sunward whistler waves, as counter propagating whistler (CPW) waves, while the predominant population is SWPWs. These SWPWs mostly occur at locations satisfying $0.1 < \Delta|B|/|B_0| \leq 0.44$ (i.e., the vertical dashed line in Figure 6f) and an inhomogeneous magnetic field. For roughly 54 percent of SWPWs, $\Delta|B|/|B_0| \geq 0.2$. On the other hand, $\Delta|B|/|B_0|$ of CPW waves are mostly below 0.2. In a CPW, the sunward wave part (pink) has a higher wave amplitude than the anti-sunward (sky blue) in the same time bin, which is clearly represented in the case related to magnetic dips. In Figure 11a in (Karbashewski et al. 2023), they show that depending on the direction of propagation obtained from Poynting vectors, five events of whistler waves are co-located with magnetic dips (as wave sources) or not (star symbols in Figures 6f and 6g). Whistler waves in this work reveal wave amplitudes peaked up to $0.06\ nT/Hz^{1/2}$ of SWPW and CPW waves, which corresponds to some black points in Figure 11a in (Karbashewski et al. 2023). The highest amplitude is shown for SWPWs in agreement with their result. However, we emphasize that this identification process specifically analyzes all whistler propagation directions in all time bins, including those in the absence of burst mode waveform data.

### 3. PROPAGATION DIRECTION OF WHISTLER WAVES

#### 3.1. Results from PSP Encounter 1

Whistler waves measured by PSP during Encounter 1 have been reported with analysis of wave parameters (Cattell et al. 2021a, 2021b; Froment et al. 2023). Before we apply the technique to all available encounters extensively, we examined $\alpha_{BV}$ for Encounter 1 and compared the results with the whistler waves collected in the burst mode and with the published results. Figure 7a shows the radial distribution of $\alpha_{BV}$, mostly near 180 degrees due to toward IMF sector structure (without crossing sector boundary) for the whole interval. However, the $\alpha_{BV}$ of a significant population (~28 percent) is between 60 and 120 degrees, meaning that the solar wind magnetic field is somewhat orthogonal to the bulk velocity (radial). As we mentioned above, electric field measurements in the case of only one component available can be underestimated for those of waves due to the topology of the solar wind magnetic field. In particular, our technique considered $\alpha_{BV}$ in order to determine propagation direction which provides a crucial threshold depending on magnetic topology (as described in Figure 5). One of the results of (Froment et al. 2023) is that 97 percent of whistler waves during Encounter 1 have a quasi-parallel WNA ($\theta \leq 45°$) relative to the background magnetic field. They presented the WNA of whistler waves with heliocentric distances and with magnetic descent levels. For later encounters, Colomban et al. (in prep.) reconstruct the lost component and confirm that the WNAs of whistler waves are predominantly quasi-parallel. Therefore, our technique can be applied reliably, at least with small angular tolerances between the solar wind magnetic field and the solar wind speed. In Figures 7b-7g, the whistler waves are also displayed according to the direction of propagation in the order of SWPWs, A-SWPWs, and CPW from the second row to the fourth row. The most important result here is that SWPWs dominate compared to A-SWPWs and CPWs, and represent 85 percent of total whistler waves during Encounter 1. ASWPWs are mostly focused between 36-44 RS, but overall, the heliocentric positions are spread between 35 and 55 RS, and the local magnetic field strength with respect to the background field is also wide. As a result, 54 percent of SWPWs occur in $\Delta|B|/|B_0|$ greater than 0.1, while ASWPWs are 39%. This result implies that inhomogeneous solar wind magnetic fields are more favorable for the generation of SWPWs than A-SWPWs. In the case of counter-propagating whistler waves, as shown in Figure 7g, events of 62% (93%) are observed in $\Delta|B|/|B_0|$ less than 0.1 (less than 0.2). Furthermore, higher wave amplitude ($B_w$) usually occurs for sunward or counter-propagating whistler waves, rare for A-SWPWs.

**3.2. Results from Encounter 1 - 11**

The data from Encounter 1 in Section 3.1 were used to verify the applicability of the technique to determine wave propagation direction on the continuously available PSP SCM and EFI spectral data using one electric field (EFI) and three magnetic field components (SCM). In the previous section, the effective length of the electric antenna has been determined for the whistler wave frequency range that allows the use of the electric field data for wave parameters processing. For later encounters, one electric field component and two magnetic field components (except the spurious component) of spectral data are available. Using all encounters of PSP with the available spectral data, we have investigated the whistler waves amplitudes and propagation direction distributions in the young solar wind. In this paper, we found and processed 12,466 whistler wave events recorded by PSP. The distributions of SWPW, A-SWPW, and counter-propagating on heliocentric distance are presented in Figure 8. The whistler occurrence rates are consistent with the previous study by Cattell et al. (2022) showing that whistler wave occurrence declines sharply below 25 RS, where only 39 whistler events (0.3%) were detected (Figure 8e). We show the strong radial dependence of the propagation direction distributions for SWPW and A-SWPW. First, we found that out of the 38 whistler waves measured closer than 25 RS, 35 (89.7%) events were SWPWs (Figures 8a and 8f). Secondly, the main populations (74%) of all the whistler waves recorded by PSP in the range of 25-50 RS had sunward propagation direction (Figure 8f). A-SWPWs and CPW waves are 19 % and 6 %, respectively. The SWPWs continue to dominate up to 40 RS (more than 80% of the recorded cases) (Figure 8d), and disappeared above 50 RS. Third, A-SWPWs occurrence rate increases with the heliocentric distance and at 40-50 RS and A-SWPWs cases were observed more frequently than SWPWs. The increase in the relative proportion of anti-sunward waves with distance is in agreement with the results of (Colomban et al. 2024) between 0.2 and 1 a.u. In the case of the CPW, it is mainly between 25 to 50 RS. Fourth, the radial distances, where the peaks of occurrence rate of SWPW, A-SWPW, and CPW were observed are shifted: 35-40 RS for SWPW, 40-45 RS for A-SWPW, and 35-40 RS for CPWs, as can be seen in Figures 8a, 8b, and 8c. The shift in occurrence rates of SWPW and A-SWPW can be caused by the waves' source localization.

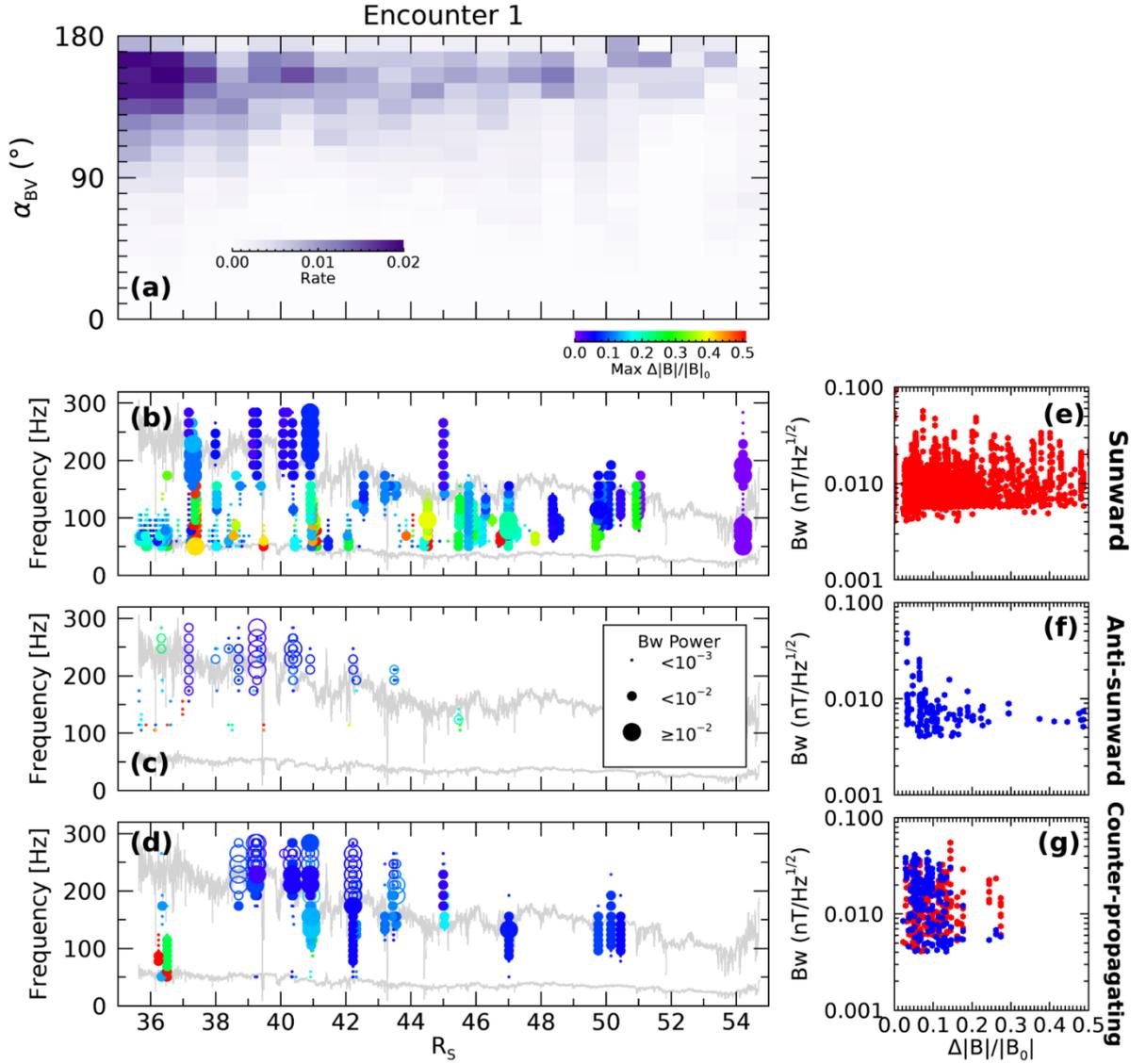

**Figure 7.** The statistical results of Encounter 1. (a) Statistics of $\alpha_{BV}$ along the radial distances. (b)-(d) Radial distribution of SWPWs (b), A-SWPWs (c), and CPWs (d) with frequency ranging from 50 to 300 Hz in the SC-frame. The gray lines are $0.1 f_{ce}$ and $f_{LH}$. The color code denotes the local magnetic field relative to the background magnetic field. The size of the circles is proportional to $\log B_w$. Open and filled circles correspond to anti-sunward and sunward propagating directions, respectively. (e)-(g) Distribution of the local magnetic field variances and the wave amplitude, corresponding to the same events in (b)-(d). The red (blue) dots represent sunward (anti-sunward) whistler waves.

23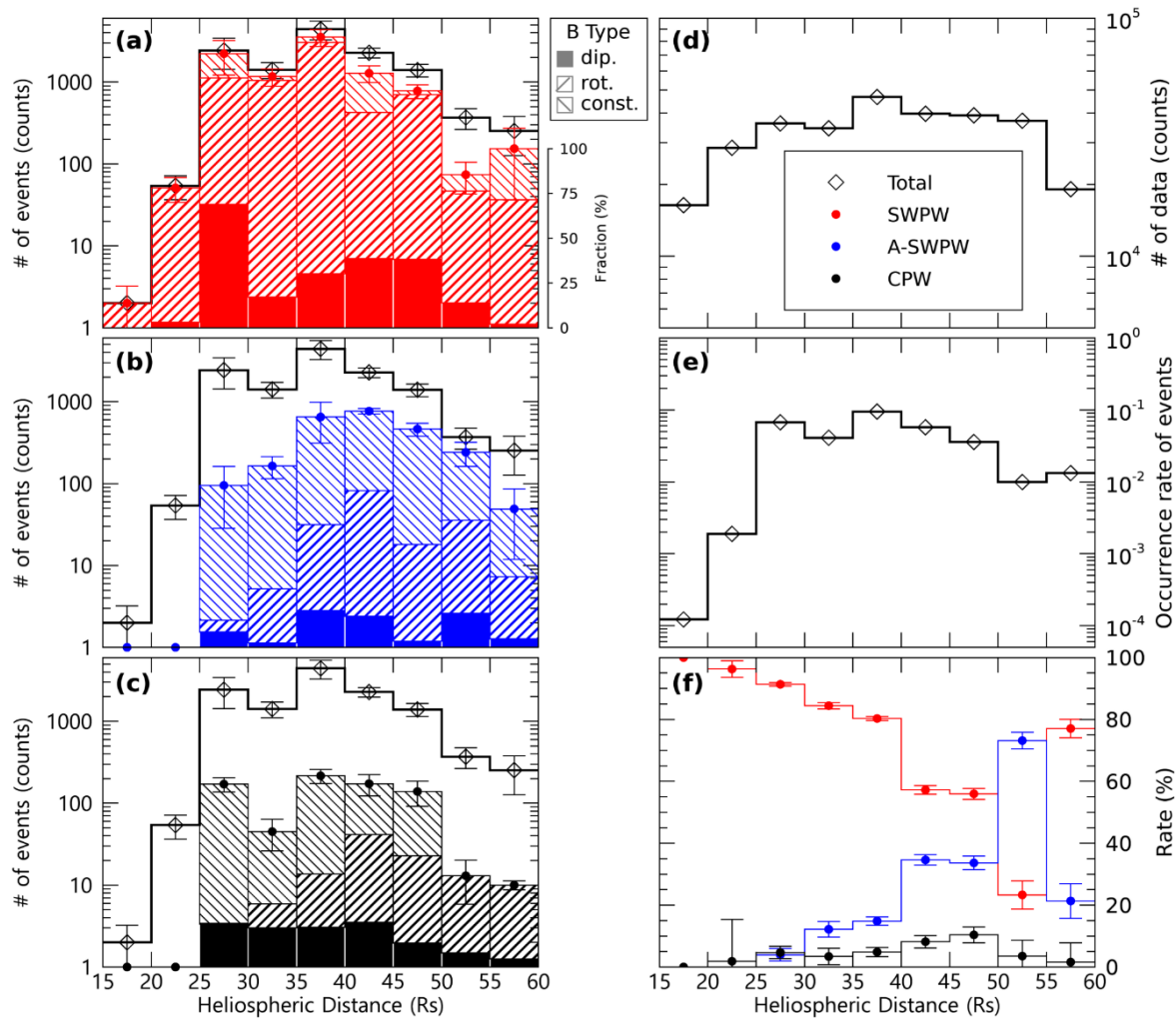

**Figure 8.** Radial distribution of whistler waves events and their propagation directions (SWPW or A-SWPW) from PSP Encounter 1-11. Panels (a)-(c) present the total number of whistler wave events (diamond symbol and black thick histogram) and each direction of wave propagation (dot symbols and colored histogram) on a logarithmic scale. Error bars show the relative variability of the number of data points within a bin. Subclass (patterns of histogram) about a population of background magnetic field inhomogeneity. The fractions of subclasses are represented on a linear scale as shown right-side axis. (a) SWPW (red dot), (b) A-SWPW (blue dot), and (c) CPW (black dot) whistler waves, respectively. (d) the BPF measurement counts, (e) the occurrence rate of all whistler waves (f) - the relative occurrence rate of three classes for wave propagating directions along the radial distance. Error bars indicate the standard errors of each bin.

Meanwhile, the notable reduction of the solar wind magnetic field and/or the large changes in the solar wind magnetic field direction have been considered as a key description of the local generation mechanism (Karbashewski et al. 2023; Froment et al. 2023). We have analyzed the perturbation of the solar wind magnetic field magnitude ($\Delta|B|/|B_0|$) and the perturbation of the solar wind magnetic field vector ($|\Delta\vec{B}|/B_0$) associated with whistler waves, following those definitions from (Karbashewski et al. 2023) to determine how propagation directions are related to such inhomogeneities. We have used $\Delta|\vec{B}|/|B_0|$ and $|\Delta\vec{B}|/B_0$ to set the threshold for each subclass: if $\Delta|\vec{B}|/|B_0|$ is greater than or equal to 0.2, it is a magnetic dip ("$B$-dips" in Figure 8), otherwise if $|\Delta\vec{B}|/B_0$ is greater than or equal to 0.4 it is a magnetic rotation ("B rots" in Figure 8), and otherwise it is constant ("B const" in Figure 8). In this part, two major differences were found for the SWPWs. First, the remarkable fraction of inhomogeneous conditions, where the occurrence of whistler B dips or B rots or both, reach 94 % for SWPWs. Most of all whistler waves (80%) were found to be associated with inhomogeneous magnetic fields, and for the A-SWPWs and CPWs only 54 % and 52 % are related to inhomogeneity. It may imply that inhomogeneous magnetic fields in the young solar wind can be favorable for generating sunward whistler waves. In comparison, for A-SWPW, the correlation with B dips is the weakest among the three types of propagating directions. For CPWs, the solar wind magnetic field condition changes with a pattern along radial distance, such that the B rots condition appearing after 30 RS increases gradually along the heliocentric distance, and the B constant condition decreases before disappearing from ~50 RS onwards (Figure 8c). We found that at all distances, SWPWs are highly correlated with inhomogeneous magnetic fields, which is consistent with the results of Karbashewski et al. (2023) and Froment et al. (2023). Also, we have found A-SWPWs association with the source region is dependent on the distance. In particular, below 35 RS, 72% of A-SWPWs are unrelated to the inhomogeneous fields.

## 4. SUMMARY AND CONCLUSION

We present a technique to determine the direction of wave propagation by making use of the continuous spectral Parker Solar Probe BPF EFI and SCM data and perform the statistics of whistler wave propagation direction based on the continuous spectral data from PSP Encounters 1-11. New findings in our work are:

1. PSP observations reveal whistler waves between 25 and 40 RS propagating predominantly toward the Sun. The abundant existence of the SWPWs supports the scenario that the source regions prevail locally in the young solar wind.

2. SWPWs are observed closer to the Sun than A-SWPWs. Below 45 RS, almost all the whistlers were found to propagate toward the Sun.

3. The direction of wave propagation is one of the key parameters of whistler waves, which determine the efficiency of interaction with electrons. Thus, the statistics of whistler wave propagation direction can be applied in order to build realistic models of wave-particle interactions efficiency.

4. Calibration of electric field measurements with the whistler waves dispersion properties showed that the effective length of the PSP electric field antennas ($L_{eff}$) critically depends on the wave frequency in the SC-frame and on the Debye length relative to the electric field antennas length.

**ACKNOWLEDGMENT**

The work of KEC, OVA, and LC were partially supported by NASA contracts 80NSSC22K0522, 80NSSC22K0433, 80NNSC19K0848, 80NSSC21K1770, and NASA's Living with a Star (LWS) program (contract 80NSSC20K0218). We thank the NASA Parker Solar Probe Mission and SWEAP team led by Justin Kasper for the use of the data. The FIELDS experiment on the Parker Solar Probe spacecraft was designed and developed under NASA contract NNN06AA01C.

**APPENDIX A:**

Figure A1 shows solar wind parameters ($N_P$ and $f_{LH}$ from magnetic field intensity) during PSP Encounters 1-11. $N_P$ varies from 10 to 5000 cm$^{-3}$ with radial distance and in different solar wind regimes as shown in Figures A1 and A2. Figure A1b shows plasma density profiles with heliocentric distance. At the same time using the proton density data, we consider the local lower hybrid frequency range seen in Figure A1c. Although we use the typical whistler wave mode frequency range from 50 Hz to 300 Hz for Encounter 1, lower hybrid frequency variation according to the local magnetic field strength and these frequencies at the perihelion of Encounter 11 reached up to 400Hz as Figure A1c shows.

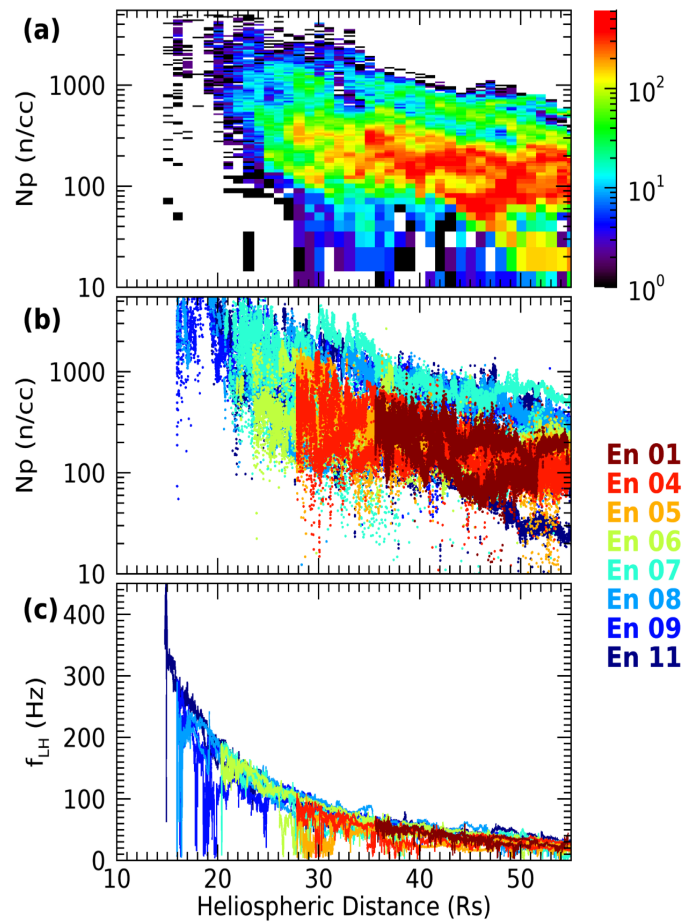

**Figure A1** Solar wind density and lower hybrid frequency profiles from Encounter 1-11. (a) Density distribution at the heliocentric distance for all encounters. Colors indicate data counts. (b) Density profiles for each encounter. (c) Variation of the lower hybrid frequency ($f_{LH}$)

according to the heliocentric distance. Each encounter in panels (b) and (c) is shown in a different color according to the legend on the right.

**REFERENCES**

Abraham, J. B., Owen, C. J., Verscharen, D., et al. 2022, ApJ, 931, 118, doi:10.3847/1538-4357/ac6605

Agapitov, O. V., Wit, T. D. de, Mozer, F. S., et al. 2020, ApJ, 891(1), L20, doi:10.3847/2041-8213/ab799c

Agapitov, O. V., Drake, J. F., Swisdak, M., et al. 2022, ApJ, 925 (The American Astronomical Society), 213

Agapitov, O. V., Drake, J. F., Swisdak, M., Choi, K.-E., & Raouafi, N. 2023, ApJL, 959 (The American Astronomical Society), L21

Artemyev, A. V., Agapitov, O. V., & Krasnoselskikh, V. V. 2013, PhPl, 20(12), 124502, doi:10.1063/1.4853615

Bale, S. D., Goetz, K., Harvey, P. R., et al. 2016, SSR, 204 (1-4), 49-82, doi:10.1007/s11214-016-0244-5

Berčič, L., Maksimović, M., Landi, S., & Matteini, L. 2019, MNRAS, 486(3), 3404-3414, doi:10.1093/mnras/stz1007

Berčič L., Verscharen, D., Owen, C. J., et al. 2021, A&A, 656, A31, doi:10.1051/0004-6361/202140970

Case, A. W., Kasper, J. C., Stevens, M. L., et al. 2020, ApJS, 246, 43, doi:10.3847/1538-4365/ab5a7b

Cattell, C., Wygant, J. R., Goetz, K., et al. 2008, GeoRL, 35, L01105, doi:10.1029/2007GL032009

Cattell, C., Short, B., Breneman, A., et al. 2021a, A&A, 650, A8, doi:10.1051/0004-6361/202039550

Cattell, C., Breneman, A., Dombeck, J., et al. 2021b, ApJL, 911(2), L29, doi:10.3847/2041-8213/abefdd

Cattell, C., Breneman, A., Dombeck, J., et al. 2022, ApJL, 924(2), L33, doi:10.3847/2041-8213/ac4015

Colomban, L., Agapitov, O. V., Krasnoselskikh, V., et al. 2023, JGRA, 128, e2023JA03142, doi:10.1029/2023JA031427

Colomban, L., Kretzschmar, M., Krasnoselskikh, V., et al. 2024, A&A, Forthcoming article, doi: 10.1051/0004-6361/202347489

Dudok de Wit, T., Krasnoselskikh, V. V., Agapitov, O., et al. 2022, JGRA, 127(4), e2021JA030018, doi:10.1029/2021JA030018

Farrell, W. M., MacDowall, R. J., Gruesbeck, J. R., Bale, S. D., & Kasper, J. C. 2020, ApJS, 249 (American Astronomical Society), 28

Farrell, W. M., Rasca, A. P., MacDowall, R. J., et al. 2021, ApJ, 915 (American Astronomical Society), 68

Feldman, W. C., Asbridge, J. R., Bame, S. J., Montgomery, M. D., & Gary, S. P. 1975, JGR, 80, 4181, doi:10.1029/JA080i031p04181

Feldman, W. C., Asbridge, J. R., Bame, S. J., Gary, S. P., & Montgomery, M. D. 1976, JGR, 81, 2377, doi:10.1029/JA081i013p02377

Feldman, W. C., J. R. Asbridge, S. J. Bame, J. T. Gosling, & D. S. Lemons 1978, JGR, 83, A11, 5297, doi:10.1029/JA083iA11p05285

Fox, N. J., Velli, M. C., Bale, S. D., et al. 2016, SSR, 204, 7, doi:10.1007/s11214-015-0211-6

Froment, C., Agapitov, O. V., Krasnoselskikh, V., et al. 2023, A&A, 672, A135, doi: 10.1051/0004-6361/202245140

Froment, C., Krasnoselskikh, V., Wit, T. D. de, et al. 2021, A&A (EDP Sciences), https://www.aanda.org/articles/aa/abs/forth/aa39806-20/aa39806-20.html

Gary, S. P., Feldman, W. C., Forslund, D. W., & Montgomery, M. D. 1975, JGR, 80, 4197

Gary, S. P., Scime, E. E., Phillips, J. L., & Feldman, W. C. 1994, JGRA, 99, 23391

Graham, G., Rae, I., Owen, C., et al. 2017, JGRA, 122, 3858, doi:10.1002/2016JA023656

Halekas, J. S., Whittlesey, P., Larson, D. E., et al. 2020, ApJS, 246(2), 22, doi: 10.3847/1538-4365/ab4cec

Halekas, J. S., Whittlesey, P., Larson, D. E., et al. 2022, ApJ, 936, 53, doi:10.3847/1538-4357/ac85b8

Jagarlamudi, V. K., Alexandrova, O., Berčič, L., et al. 2020, ApJ, 897, 118, doi:10.3847/1538-4357/ab94a1

Jagarlamudi, V. K., Dudok de Wit, T., Froment, C., et al. 2021, A&A, 650, A9, doi:10.1051/0004-6361/202039808

Kajdič, P., Alexandrova, O., Maksimovic, M., Lacombe, C., & Fazakerley, A. N. 2016, ApJ., 833, 172, doi:10.3847/1538-4357/833/2/172

Karbashewski, S., Agapitov, O. V., Kim, H. Y., et al. 2023, ApJ, 947(2), 73. Doi:10.3847/1538-4357/acc527

Kasper, J. C., Abiad, R., Austin, G., et al. 2016, SSRv, 204, 131, doi:10.1007/s11214-015-0206-3

Krasnoselskikh, V., Larosa, A., Agapitov, O., et al. 2020, ApJ, 893 (American Astronomical Society), 93

Kretzschmar, M., Chust, T., Krasnoselskikh, V., et al. 2021, A&A, 656, A24, doi:10.1051/0004-6361/202140945

Larosa, A., Krasnoselskikh, V., Wit, T. D. de, et al. 2021, A&A, 650 (EDP Sciences), A3

Lee, S.-Y., Lee, E., Seough, J., et al. 2018, JGRA, 123, 3277–3290, doi:10.1029/2017JA024960

Lee, S.-Y., Lee, E., & Yoon, P. 2019, ApJ, 876, 117, doi:10.3847/1538-4357/ab12db

Lemons, D. S., & Feldman, W. C. 1983, JGR, 88, 6881, doi:10.1029/JA088iA09p06881

López, R. A., Lazar M., Shaaban S. M., et al. 2019, ApJL 873 L20, doi: 10.3847/2041-8213/ab0c95

Maksimovic, M., Zouganelis, I., Chaufray, J. Y., et al. 2005, JGRA, 110, A09104, doi:10.1029/2005JA011119

Malaspina, D. M., Ergun, R. E., Bolton, M., et al. 2016, JGRA, 121, 5088, doi: 10.1002/2016JA022344

Malaspina, D. M., Halekas, J., Berčič, L., et al. 2020, ApJS, 246(2), 21, doi:10.3847/1538-4365/ab4c3b

Mozer, F. S., Agapitov, O. V., Bale, S. D., et al. 2020, ApJS, 246(2), 68, doi:10.3847/1538-4365/ab7196

Moullard, O., Burgess, D., Salem, C., Mangeney, A., Larson, D. E., & Bale, S. D. 2001, JGR, 106, A5, doi:10.1029/2000JA900144

Pagel, C., Gary, S. P., de Koning, C. A., Skoug, R. M., & Steinberg, J. T. 2007, JGRA, 112, A04103

Pierrard, V., Lazar, M., & Schlickeiser, R. 2011, SoPh, 269(2), 421–438, doi:10.1007/s11207-010-9700-7

Pilipp, W. G., Miggenrieder H., Montgomery M. D. et al. 1987, JGR, 92 1075, doi: 10.1029/JA092iA02p01075

Raouafi, N. E., Matteini, L., Squire, J., et al. 2023, SSR, 219, 8, doi:10.1007/s11214-023-00952-4

Roberg-Clark, G. T., Drake, J. F., Swisdak, M., & Reynolds, C. S. 2018, ApJ, 867, 154, doi:10.3847/1538-4357/aae393

Roberg-Clark, G. T., Agapitov, O., Drake, J. F., Pagel & Swisdak, M. 2019, ApJ, 887(2), 190, doi:10.3847/1538-4357/ab5114

Romeo, O. M., Braga, C. R., Badman, S. T., et al. 2023, ApJ, 954, 2, doi:10.3847/1538-4357/ace62e

Rosenbauer, H., Schwenn R., Marsch E., et al. 1977, JGZG, 42 561
Saito, S., & Gary, S. P. 2007, JGRA, 112, doi:10.1029/2006JA012216

Saito, S., & Gary, S. P. 2007, JGRA, 112, doi:10.1029/2006JA012216

Salem, C. S., Pulupa, M., Bale, S. D., & Verscharen, D. 2023, A&A, 675, A162, doi:10.1051/0004-6361/202141816

Seough, J., Nariyuki, Y., Yoon, P. H., & Saito, S. 2015, ApJ, 811(1), 5, doi:10.1088/2041-8205/811/1/L7

Shaaban, S. M., Lazar, M., Yoon, P. H., Poedts, S., & López, R. A. 2019, MNRAS, 486(4), 4498–4507, doi:10.1093/mnras/stz830

Stansby, D., Horbury, T. S., Chen, C. H. K., & Matteini, L. 2016, ApJ, 829, L16, doi: 10.3847/2041-8205/829/1/L16

Steinvall, K., Khotyaintsev, Y. V., Cozzani, G., et al. 2021, A&A, 656, A9, doi:10.1051/0004-6361/20214085

Štverák, Š., Travnicek, P., Maksimovic, M., Marsch, E., Fazakerley, A. N., & Scime, E. E. 2008, JGRA, 113, A3, doi:10.1029/2007JA012733

Štverák, Š., Maksimovic, M., Travnicek, P. M., et al. 2009, JGR, 114, A05104, doi:10.1029/2008JA013883

Tang, B., Zank, G. P., & Kolobov, V. I. 2020, ApJ, 892(2), 95. doi: 10.3847/1538-4357/ab7a93

Tong, Y., Vasko, I. Y., Artemyev, A. V., Bale, S. D., & Mozer, F. S. 2019, ApJ, 878, 41, doi: /10.3847/1538-4357/ab1f05

Vasko, I. Y., Kuzichev, I. V., Artemyev, A. V., et al. 2020, PhPl, 27(8), 082902, doi:10.1063/5.0003401

Vo, T., Lysak, R., & Cattell, C. 2022, PhPl, 29(1), 012904, doi:10.1063/5.0074474

Vocks, C. 2012, SSRv, 172, 303, 314, doi:10.1007/s11214-011-9749-0

Vocks, C., & Mann, G. 2003, ApJ, 593 (2), 1134–1145, doi:10.1086/376682

Vocks, C., Salem, C., Lin, R. P., & Mann, G. 2005, ApJ, 627, 540, doi:746 10.1086/430119

Wilson III, L. B., Cattell, C. A., Kellogg, P. J., Wygant, J. R., Goetz, K., Breneman, A., Kersten, K. 2011, GRL, 38, 17, doi:

10.1029/2011GL048671

Wilson III, L. B., Chen, L.-J., Wang, S., Schwartz, S. J., Turner, D. L., Stevens, M. L., Kasper, J. C., Osmane, A., Caprioli, D., Bale, S. D., Pulupa, M. P., Salem, C. S., & Goodrich, K. A. 2019, ApJS, 243, 8, doi:10.3847/1538-4365/ab22bd

Wilson, III, L. B., Koval, A., Szabo, A., Breneman, A., Cattell, C. A., Goetz, K., et al. 2013, JGRA, 118(1), 5–16, doi:10.1029/2012JA018167

Zenteno-Quinteros B. & Moya P. S. 2022, FrPhy, 10, 910193, doi:10.3389/fphy.2022.910193